\documentclass[prd,twocolumn,preprint numbers, aps,superscriptaddress,nofootinbib,showpacs]{revtex4-1}
 \usepackage{amsmath}
\usepackage{amssymb}
\usepackage{graphicx}
\usepackage{color}
\usepackage{bm}
\usepackage{times}
\usepackage{hyperref}
\hypersetup{
  colorlinks=true,
  citecolor=blue,
  linkcolor=blue,
  urlcolor=blue}

\usepackage{epsfig}% Include figure files
\usepackage{dcolumn}% Align table columns on decimal point
\usepackage{float}

\usepackage{epsfig}% Include figure files
\usepackage{dcolumn}% Align table columns on decimal point
\usepackage[normalem]{ulem}
\def\be{\begin{equation}}
\def\ee{\end{equation}}
\def\bea{\begin{eqnarray}}
\def\eea{\end{eqnarray}}
\def\gsim{\ \rlap{\raise 2pt\hbox{$>$}}{\lower 2pt \hbox{$\sim$}}\ }
\def\lsim{\ \rlap{\raise 2pt\hbox{$<$}}{\lower 2pt \hbox{$\sim$}}\ }
\def\dslash{\kern-4pt \not{\hbox{\kern-2pt $\partial$}}}
\def\pslash{\not{\hbox{\kern-2pt p}}}

\begin{document}

\DeclareGraphicsExtensions{.eps,.ps}

\title{Analysis of four-zero textures in $3+1$ framework}

\author{Debasish Borah}
\email{dborah@iitg.ernet.in}
\affiliation{Department of Physics, Indian Institute of Technology Guwahati, Assam-781039, India}

\author{Monojit Ghosh}
\email{monojit@tmu.ac.jp}
\affiliation{Physical Research Laboratory, Navrangpura, Ahmedabad 380009, India}
\affiliation{Department of Physics, Tokyo Metropolitan University, Hachioji, Tokyo 192-0397, Japan}

\author{Shivani Gupta}
\email{shivani.gupta@adelaide.edu.au}
\affiliation{
Center of Excellence in Particle Physics (CoEPP), University of Adelaide, Adelaide SA 5005, Australia}

\author{Suprabh Prakash}
\email{prakash3@mail.sysu.edu.cn}
\affiliation{
School of Physics, Sun Yat-Sen (Zhongshan) University, Guangzhou 510275, P.~R.~China}

\author{Sushant K. Raut}
\email{raut@kth.se}
\affiliation{
Department of Theoretical Physics, School of Engineering Sciences, KTH Royal Institute of Technology, AlbaNova University Center, 106 91 Stockholm, Sweden}

%\maketitle
\begin{abstract}
The presence of a zero texture in the neutrino mass matrix can indicate the 
presence of an underlying symmetry 
%underlying the new physics 
which can generate neutrino mass and mixing.  
In this paper, for the first time we study the four-zero textures of the
low energy neutrino mass matrix in the presence of an extra light-sterile
neutrino i.e., the 3+1 neutrino scheme.
In our analysis we find that out of the 210 possible four-zero textures
only 15 textures are allowed. We divide the allowed four-zero textures into 
two classes -- class $A$ in which
the value of mass matrix element $M_{ee}$ is zero and class $B$ in which
$M_{ee}$ is non-zero. In this way we obtain ten
possible four-zero textures in class $A$ and five possible
four-zero textures in class $B$. 
In our analysis we find that, for
normal hierarchy the allowed number of textures in class $A$ ($B$) is nine (three).
For the case of inverted hierarchy we find that, two textures in class $A$ 
are disallowed and these textures are different from the disallowed
textures for normal hierarchy in class $A$. However, we find that
all the five textures in class $B$ are allowed for the inverted hierarchy. 
Based on analytic expressions for the elements $M_{\alpha\beta}$, we 
discuss the reasons for certain textures being disallowed. We also
discuss the correlations between the different parameters of the allowed textures. 
Finally, we present the implications of our study on experimental searches 
for neutrinoless double beta decay.

\end{abstract}

\pacs{12.60.-i,12.60.Cn,14.60.Pq}
\preprint{ADP-16 - 20 / T975}
\maketitle

\section{Introduction}
\label{sec:intro}

Non-zero neutrino masses and large leptonic mixing have been reported 
and confirmed by 
several experiments~\cite{PDG,PDG1,PDG2,PDG3,PDG4,kamland} over the last twenty years. 
The Standard Model (SM) of particle physics whose last missing piece, 
the Higgs boson was discovered at the Large Hadron Collider (LHC) in 2012, 
fails to explain the non-zero neutrino masses and mixing due to the absence of 
right-handed neutrinos, thereby forbidding a renormalizable mass term similar to the charged fermions. 
Consequently, there have been overwhelming experimental and theoretical efforts 
in the particle physics community 
to understand the new physics behind it. 
Even if the SM is extended by including three right-handed neutrinos and hence allowing Dirac mass terms, 
the corresponding Dirac Yukawa couplings have to be fine tuned to the level of $10^{-12}$ in order to 
generate sub-eV neutrino masses. This led to the adoption of the popular seesaw mechanism for generating tiny 
neutrino masses. 
In this mechanism, neutrino masses originate from the seesaw between the electroweak scale 
and the scale of additional heavy particles introduced.
Seesaw models can be broadly divided into three types: 
type I~\cite{ti,ti1,ti2,ti3,ti4}, type II~\cite{tii,tii1,tii2,tii3,tii4,tii5,tii6} and type III~\cite{tiii}. 
All these models can successfully explain the sub-eV scale neutrino masses and their mixings which have been confirmed again 
by the recent experiments MINOS~\cite{minos}, T2K~\cite{T2K}, NO$\nu$A~\cite{nova}, 
Double ChooZ~\cite{chooz}, Daya-Bay~\cite{daya} and RENO~\cite{reno}. 
Various global fits of the world data to neutrino oscillation parameters have 
provided us with their best-fit values and $3\sigma$ allowed 
ranges~\cite{schwetz14,valle14,Capozzi:2013csa}.
%are shown in table~\ref{tab:data1} and~\ref{tab:data2} respectively. 
%In table~\ref{tab:data1} and~\ref{tab:data2}, $\theta_{ij}$ are the three neutrino mixing angles, $\Delta m_{ij}^2=m_i^2-m_j^2$ are the two mass-squared 
%differences and $\delta$ is the leptonic Dirac CP phase. Although $\delta$ can take any values as seen from the above $3\sigma$ data, 
%recent results from T2K and NO$\nu$A have favoured its value to be $-\pi/2$~\cite{diracphase,diracphasenova}.

If neutrinos are Majorana fermions whose masses originate from conventional seesaw mechanisms, 
then two Majorana CP phases also appear in the mixing matrix. However, 
they do not affect neutrino oscillation probabilities and hence remain 
undetermined at neutrino oscillation experiments. 
Apart from the Majorana CP phases, the absolute neutrino mass is 
also unknown as the experiments can measure only the two mass-squared 
differences. We however, 
have an upper bound on the lightest neutrino mass from the Planck data in terms of the sum of absolute 
neutrino masses $\sum_i \lvert m_i \rvert < 0.23$ eV~\cite{Planck15}. 
The neutrino parameters like absolute neutrino mass and Majorana CP phases which remain undetermined at 
neutrino oscillation experiments can however, have interesting
consequences at neutrinoless double beta decay 
(NDBD) experiments like KamLAND-Zen~\cite{kamland_zen} and GERDA~\cite{GERDA} based on Xenon-136 and Germanium-76 nuclei, 
respectively. 
%\sout{To correlate any future experimental signatures with the underlying theory, 
%one must have guidelines from the model building point of view regarding
% the values of these neutrino parameters. }

Apart from the three sub-eV scale active neutrinos, 
some experiments also suggest the presence of additional light
sterile neutrinos at the eV scale\footnote{For a review, see Ref.~\cite{whitepaper}.}. Data from the nine year Wilkinson
Microwave Anisotropy Probe (WMAP)  
point towards the existence of additional light degrees of freedom $N_{\text{eff}} = 3.84 \pm 0.40$~\cite{wmap9}.
But the recent Planck data show that it is possible to have one
extra light sterile neutrino in the eV scale only
if one deviates from the standard $\Lambda$CDM model~\cite{Planck15}.
However, the issue of the existence of light sterile neutrinos is 
not yet settled due to anomalies found in accelerator and reactor based neutrino experiments. 
The LSND accelerator experiment saw anomalies in the anti-neutrino 
flux~\cite{LSND1} that could not be explained with the three-neutrino 
oscillation picture. Subsequently the antineutrino results from MiniBooNE~\cite{miniboone} 
also supported the LSND findings. 
Similar anomalies have also been observed at reactor neutrino
experiments~\cite{react} as well as gallium solar neutrino experiments~\cite{gall1,gall2}. 
Although the Planck results do not favor such light sterile neutrinos,
there could be unknown non-standard cosmology 
behind the existence of such relativistic degrees of freedom which do not
show up in cosmological observations. 
Some interesting discussions on light sterile neutrinos from the point of 
view of cosmology as well as oscillation experiments can be found
in Ref.~\cite{cosmo_ste1,cosmo_ste2,cosmo_ste3,cosmo_ste4} and references therein. 
This has generated a new challenge and some activities in order to
develop a particle physics model to accommodate light sterile neutrinos
and their mixing with active neutrinos, 
as well as to have a consistent cosmological model. 

% \begin{table}
% \begin{center}
% \begin{tabular}{|c|c|c|}
% \hline
% Parameters & Normal Hierarchy (NH) & Inverted Hierarchy (IH) \\
% \hline
% $ \frac{\Delta m_{21}^2}{10^{-5} \text{eV}^2}$ & $7.02-8.09$ & $7.02-8.09 $ \\
% $ \frac{|\Delta m_{31}^2|}{10^{-3} \text{eV}^2}$ & $2.317-2.607$ & $2.307-2.590 $ \\
% $ \sin^2\theta_{12} $ &  $0.270-0.344 $ & $0.270-0.344 $ \\
% $ \sin^2\theta_{23} $ & $0.382-0.643$ &  $0.389-0.644 $ \\
% $\sin^2\theta_{13} $ & $0.0186-0.0250$ & $0.0188-0.0251 $ \\
% $ \delta $ & $0-2\pi$ & $0-2\pi$ \\
% \hline
% \end{tabular}
% \end{center}
% \caption{Global fit $3\sigma$ values of neutrino oscillation parameters~\cite{schwetz14}}
% \label{tab:data1}
% \end{table}
% 
% \begin{table}
% \begin{center}
% \begin{tabular}{|c|c|c|}
% \hline
% Parameters & Normal Hierarchy (NH) & Inverted Hierarchy (IH) \\
% \hline
% $ \frac{\Delta m_{21}^2}{10^{-5} \text{eV}^2}$ & $7.11-8.18$ & $7.11-8.18 $ \\
% $ \frac{|\Delta m_{31}^2|}{10^{-3} \text{eV}^2}$ & $2.30-2.65$ & $2.20-2.54 $ \\
% $ \sin^2\theta_{12} $ &  $0.278-0.375 $ & $0.278-0.375 $ \\
% $ \sin^2\theta_{23} $ & $0.393-0.643$ &  $0.403-0.640 $ \\
% $\sin^2\theta_{13} $ & $0.0190-0.0262$ & $0.0193-0.0265 $ \\
% $ \delta $ & $0-2\pi$ & $0-2\pi$ \\
% \hline
% \end{tabular}
% \end{center}
% \caption{Global fit $3\sigma$ values of neutrino oscillation parameters~\cite{valle14}}
% \label{tab:data2}
% \end{table}

In a three active and one sterile neutrino framework, the light neutrino
mass matrix is a $4\times 4$ complex symmetric matrix, 
assuming the neutrinos to be Majorana particles. Irrespective of the
dynamical origin of such a mass matrix, this can be parametrized by four
masses, six angles and six phases, a total of sixteen parameters. Since
many of these parameters are not accurately determined by experiments, 
one can consider them to be free parameters. However, if the underlying
symmetry of the theory is such that it relates some of these parameters or 
fixes them to some numerical values, then the model becomes more
predictive and can be tested in the ongoing experiments. 
One such scenario is the zero texture models where some of the
elements in the leptonic mass matrices are zero. 
For a survey of such zero textures in lepton mass matrices, we refer to the recent article Ref.~\cite{Ludl2014}. 
Different possible flavor symmetries can be responsible for such zero
textures in the mass matrices~\cite{Joshipura:2016hvn,Karmakar:2014dva,texturesym,texturesym1,texturesym2,texturesym3,texturesym4,texturesym5,texturesym6,texturesym7,texturesym8,texturesym9} 
within the framework of different seesaw models. 
Several earlier studies related to zero textures in the three neutrino picture can be found in 
Refs.~\cite{Xing:2004ik,onezero,onezero1,onezero2,onezero3,alltex,texturesym9,twozero,twozero1,twozero2,twozero3,twozero4,twozero5,twozero6,twozero7,twozero8,twozero9,twozero10,twozero11}. 
Recently, the possibilities of such zero textures were explored in the case of three
active and one sterile neutrino (the $3+1$ framework) as well~\cite{3+12zero, 3+11zero, 3+13zero, 3+1zero}. 
The possibilities of one-zero, two-zero, three-zero textures have already been explored in these works. 
The authors of these works have pointed out the allowed zero texture
mass matrices from the available data of the mixing angles and mass squared differences. 
In this work, for the first time, we study the possibility of four-zero textures in 
the $4\times 4$ mass matrix of the $3+1$ neutrino scenario.

The paper is organized as follows. In Section~\ref{sec2}, we discuss
the parametrization of the $4\times4$ low energy neutrino mass matrix in the 3+1
scenario and the methodology
that we adopt to obtain the viable textures. In Section~\ref{res}, we
present our numerical results along with the analytical explanations.
In Section~\ref{symmetry}, we discuss the origin of the four-zero texture
via flavor symmetries and finally we summarize our results in Section~\ref{sum}. \\

\section{Four-zero textures in $3+1$ scenario}
\label{sec2}

For our analysis we choose a basis where the charge lepton mass matrix is diagonal. 
Therefore the lepton mixing matrix is simply $U_\nu = U_{PMNS} = U$.
Hence any complex symmetric $4\times4$ light neutrino mass matrix can be written as 
\begin{equation}
M_{\nu} = U M^{\text{diag}}_{\nu} U^T,
\label{mnu}
\end{equation}
where $M^{\text{diag}}_{\nu} = \text{diag}(m_1, m_2, m_3, m_4)$ is the
 diagonal form of the light neutrino mass matrix. 
As already mentioned, the diagonalizing matrix $U$ is the $4\times4$ version of the
 Pontecorvo-Maki-Nakagawa-Sakata (PMNS) leptonic mixing matrix which can be written as 
$$ U = V P,$$
where $V$ is the mixing matrix for Dirac neutrinos containing six angles
 and three Dirac phases and $P$ is a diagonal matrix containing three
  Majorana phases. The matrix $V$ can be parametrized as~\cite{param3+1}
\begin{equation}
V = R_{34} \tilde{R}_{24}\tilde{R}_{14}R_{23}\tilde{R}_{13} R_{12},
\end{equation}
%\begin{widetext}
where the rotation matrices $R, \tilde{R}$ can be further parametrized as (for example $R_{34}$ and $\tilde{R}_{14}$)
\begin{eqnarray}
R_{34}&=&\begin{pmatrix}
1 & 0 & 0 & 0\\
0 & 1 & 0 & 0 \\
0 & 0 & c_{34} & s_{34} \\
0 & 0 & -s_{34} & c_{34}
\end{pmatrix}, \\
\tilde{R}_{14}&=&\begin{pmatrix}
c_{14} & 0 & 0 & s_{14} e^{-i\delta_{14}}\\
0 & 1 & 0 & 0 \\
0 & 0 & 1 & 0 \\
-s_{14} e^{i\delta_{14}} & 0 & 0 & c_{14} 
\end{pmatrix},
\end{eqnarray}
where $c_{ij} = \cos{\theta_{ij}}, \; s_{ij} = \sin{\theta_{ij}}$ and $\delta_{ij}$ are the Dirac CP phases. 
The diagonal phase matrix is given by $P = \text{diag} (1, e^{-i\alpha/2}, e^{-i(\beta/2 - \delta_{13})}, e^{-i(\gamma/2 - \delta_{14})})$ 
which contains three Majorana phases. 
In our choice of parametrization, all the phases can vary from $-\pi$ to $\pi$.
For normal hierarchy (i.e., NH: $m_3 > m_2 > m_1$), the light
 neutrino mass matrix in the mass basis can be written as 
 %\red{Usually one always writes $\Delta m^2_{ij}$ with the convention that $i>j$. Should we rewrite 
 %these two equations with $\Delta m_{32}^2$ instead of $\Delta m_{23}^2$?}
\begin{widetext}
\begin{equation}
M^{\text{diag}}_{\nu} 
= \text{diag}(m_1, \sqrt{m^2_1+\Delta m_{21}^2}, \sqrt{m_1^2+\Delta m_{31}^2}, \sqrt{m_1^2+\Delta m_{41}^2}),
\end{equation}
where $\Delta m^2_{ij} = m_i^2 - m_j^2$.
Similarly for inverted hierarchy (i.e., IH: $m_2 > m_1 > m_3$), the diagonal mass matrix is
\begin{equation}
M^{\text{diag}}_{\nu} = \text{diag}(\sqrt{m_3^2-\Delta m_{32}^2-\Delta m_{21}^2}, \sqrt{m_3^2-\Delta m_{32}^2}, m_3, \sqrt{m_3^2+\Delta m_{43}^2}).
\end{equation}
\end{widetext}

If neutrinos are Majorana fermions as predicted by the conventional seesaw mechanisms, 
then the $4\times 4$ neutrino mass matrix in the $3+1$ neutrino scenario is
 complex symmetric and hence has ten independent complex elements. 
If $n$ number of elements among them are assumed to be zero then the total
 number of structurally different Majorana neutrino mass matrices with $n$-zero texture is
\begin{equation}
\label{prmtn}
^{10}C_n=\frac{10!}{n!(10-n)!}. 
\end{equation}
Thus for $n=4$, there are in total 210 possible four-zero textures.
 But out of them, 195 textures can be ruled by the following argument.
In the previous work on two-zero textures in the $3+1$ neutrino framework~\cite{3+12zero}, 
it was shown that the simultaneous existence of zeros
 in active and extended sterile sector is phenomenologically disallowed.
% The magnitude of the sterile mass term is much larger
%than active neutrino mass term and thus, there is no cancellation
%between active and sterile neutrino contributions.
Thus two-zero textures are only possible in the first $3\times3$ block of the $4\times4$ mass matrix. 
Since two-zero textures are only a subset of four-zero textures, we only need
 to consider the possibility of having all the four zeros in the $3\times3$ block of the $4\times4$ mass matrix.
This rules out 195 of the possible four-zero textures.
Considering all the four-zero textures to be in the first $3\times3$
 block of the $4\times4$ neutrino mass matrix, 
the total number of independent texture zero mass matrices are
 $^6C_4 = 15$. We divide them into two classes: four-zero textures with
 $M_{ee}=0$ (class $A$) and four-zero textures
with $M_{ee} \neq 0$ (class $B$). The 15 possible texture zero matrices 
are listed below:
\begin{widetext}
\begin{eqnarray}
A_1 &:&\begin{pmatrix}
0& 0 &\times & \times\\
 0 & 0 &\times & \times \\
\times& \times & 0 &\times \\
\times & \times & \times & \times 
\end{pmatrix} , 
  A_2 :\begin{pmatrix}
0& \times &0 & \times\\
\times& 0 &\times & \times \\
0 & \times & 0 &\times \\
\times & \times & \times & \times 
\end{pmatrix},
  A_3 :\begin{pmatrix}
0& \times &\times & \times\\
\times& 0 & 0 & \times \\
\times& 0 & 0 &\times \\
\times & \times & \times & \times 
\end{pmatrix};  \\ \nonumber
A_4 &:&\begin{pmatrix}
0& 0 &0 & \times\\
0 & 0 &\times & \times \\
 0 & \times & \times &\times \\
\times & \times & \times & \times 
\end{pmatrix} , 
  A_5 :\begin{pmatrix}
0& 0 &\times & \times\\
0 & 0 & 0 & \times \\
\times& 0 & \times &\times \\
\times & \times & \times & \times 
\end{pmatrix},
  A_6 :\begin{pmatrix}
0& \times &0 & \times\\
\times & 0 & 0 & \times \\
0 & 0 & \times &\times \\
\times & \times & \times & \times 
\end{pmatrix}, \\ \nonumber
A_7 &:&\begin{pmatrix}
0& 0 &0 & \times\\
0 & \times &\times & \times \\
 0 & \times & 0 &\times \\
\times & \times & \times & \times 
\end{pmatrix} , 
  A_8 :\begin{pmatrix}
0& 0 &\times & \times\\
0 & \times & 0 & \times \\
\times& 0 & 0 &\times \\
\times & \times & \times & \times 
\end{pmatrix},
  A_9 :\begin{pmatrix}
0& \times &0 & \times\\
\times & \times & 0 & \times \\
0 & 0 & 0 &\times \\
\times & \times & \times & \times 
\end{pmatrix}, \\ \nonumber
 A_{10}&:&\begin{pmatrix}
0 & 0 & 0 & \times \\
0 & \times &0 & \times \\
 0 & 0 & \times &\times \\
\times & \times & \times & \times
\end{pmatrix},
 B_1 :\begin{pmatrix}
\times& 0 &\times & \times\\
0 & 0 & 0 & \times \\
\times& 0 & 0 &\times \\
\times & \times & \times & \times 
\end{pmatrix},
  B_2 :\begin{pmatrix}
\times& \times &0 & \times\\
\times & 0 & 0 & \times \\
0 & 0 & 0 &\times \\
\times & \times & \times & \times 
\end{pmatrix}, \\ \nonumber
 B_3 &:&\begin{pmatrix}
\times & 0 & 0 & \times \\
0 & 0 &\times & \times \\
 0 & \times & 0 &\times \\
\times & \times & \times & \times  
\end{pmatrix} ,  
  B_4 :\begin{pmatrix}
\times & 0 & 0 & \times \\
0 & 0 &0 & \times \\
 0 & 0 & \times &\times \\
\times & \times & \times & \times 
\end{pmatrix},
  B_5 :\begin{pmatrix}
\times & 0 & 0 & \times \\
0 & \times &0 & \times \\
 0 & 0 & 0 &\times \\
\times & \times & \times & \times 
\end{pmatrix}.
\end{eqnarray}
\end{widetext}
We see that there are ten four-zero textures in class $A$ and five four-zero textures in class $B$.
Using the parametrization mentioned above, the elements of the $4\times 4$ complex symmetric 
light neutrino mass matrix (cf. Eq.~\ref{mnu}) can be written in terms of sixteen parameters
\footnote{The full expressions of all the elements are given in Appendix~\ref{appen1}.}.
These sixteen parameters are: four mass eigenvalues i.e., $m_1$, $m_2$, $m_3$, $m_4$, 
six mixing angles i.e., $\theta_{12}$, $\theta_{23}$, $\theta_{13}$, $\theta_{14}$, $\theta_{24}$, $\theta_{34}$ and 
six phases i.e., $\delta_{13}$, $\delta_{14}$, $\delta_{24}$, $\alpha$, $\beta$, $\gamma$.

The four-zero texture condition of Eq.~\ref{mnu} can be written as,
\begin{eqnarray}
 a_i m_1 + b_i m_2 + c_i m_3 + d_i m_4 = 0, \quad (i \in \{1-4\})
\end{eqnarray}
which is a system of four non-linear complex algebraic equations. To solve this, we proceed in the following way. We
decompose the four complex equations into eight real equations
 by setting the real part and imaginary part individually to be zero i.e.,
 \begin{eqnarray}
 && a_i^\prime m_1 + b_i^\prime \sqrt{m^2_1+\Delta m_{21}^2} \\  \nonumber
 &+& c_i^\prime \sqrt{m^2_1+\Delta m_{31}^2} + d_i^\prime \sqrt{m^2_1+\Delta m_{41}^2} = 0 ~ {\rm (for ~ NH)}~, \label{ih} \\
 && a_i^\prime \sqrt{m_3^2+\Delta m_{32}^2-\Delta m_{21}^2}  + \\ \nonumber && b_i^\prime \sqrt{m_3^2+\Delta m_{32}^2}
 + c_i^\prime m_3 + d_i^\prime \sqrt{m_3^2+\Delta m_{43}^2} = 0 ~ {\rm (for ~ IH)}~, 
 \end{eqnarray}
where $i \in \{1-8\}$. Now we have eight equations and sixteen variables. In order to solve it,
 we supply the input values of the eight parameters $\Delta m_{21}^2$, $\Delta m_{31}^2$ ($\Delta m_{32}^2)$,
$\Delta m_{41}^2$ ($\Delta m_{43}^2$), $\theta_{12}$, $\theta_{23}$, $\theta_{13}$, $\theta_{14}$, $\theta_{24}$ 
randomly within their allowed ranges and solve for the remaining eight parameters
i.e., $m_1$ ($m_3$), $\theta_{34}$ and the six phases for NH (IH) using the damped Newton-Raphson method. 
If the solutions satisfy the condition $m_1$ ($m_3) > 0$ and
 $\theta_{34}$ is within its allowed values, then that texture will be allowed in NH (IH).
The allowed ranges used for the active neutrino parameters are consistent with the $3 \sigma$ ranges of the present global fits~\cite{schwetz14,valle14,Capozzi:2013csa}.
For the sterile neutrino parameters, we vary $\theta_{14}$ between
 $0^\circ$ to $20^\circ$, $\theta_{24}$ between $0^\circ$ to $11.5^\circ$
  and $\Delta m_{41}^2$ ($\Delta m_{43}^2$) between (0.7 - 2.5) eV$^2$. The allowed range of
$\theta_{34}$ is taken as $0^\circ < \theta_{34} < 30^\circ$~\cite{globalfit}. The constraints on $\theta_{14}$ can be drawn 
from Ref.~\cite{An:2014bik}. 
For $\theta_{24}$ and $\theta_{34}$, one can refer to Ref.~\cite{Adamson:2011ku, jones_icecube2016}. 
These experiments exclude a part of the $\sin^2\theta_{i4}(i=1,2,3)-\Delta m^2_{41}$ 
parameter space with which their data is not compatible. As a result, 
they put only an upper limit on the mixing angles (correlated with 
mass-squared difference) because they analyse stand-alone data. 
Ref.~\cite{globalfit} does a global analysis of the available data and provides 
constraints on the active-sterile mixings where the lower allowed limit is greater than zero. 
In all these works, the results have been quoted at different confidence levels. 
To follow a conservative approach, we consider the upper allowed 
$\sim3\sigma$ limits of the active-sterile mixing angles from Ref.~\cite{globalfit}. 
However, we put their lower allowed limit to be 0.

\begin{table}
\begin{center}
\begin{tabular}{|c|c|c|}
\hline
Possibility & NH & IH \\
\hline
Allowed in class A & $A_1$, $A_2$, $A_4$, $A_5$, $A_6$,  & $A_1$, $A_2$, $A_3$, $A_4$,  \\
                   & $A_7$, $A_8$, $A_9$, $A_{10}$         & $A_5$, $A_6$, $A_7$, $A_8$ \\
Disallowed in class A & $A_3$ & $A_9$, $A_{10}$ \\
Allowed in class B & $B_3$, $B_4$, $B_5$ & $B_1$, $B_2$, $B_3$, $B_4$, $B_5$  \\
Disallowed in class B & $B_1$, $B_2$ & -- \\
\hline
\end{tabular}
\end{center}
\caption{The allowed and disallowed four-zero textures}
\label{su}
\end{table}

\section{Results}
\label{res}

In this section we present our numerical results and try to explain those results from the analytic expressions 
of the mass matrix elements. In table~\ref{su}, we summarize our results
in terms of allowed and disallowed textures. Below we discuss them in detail.

\subsection{Allowed textures in class $A$}

\begin{figure*}
\begin{center}
\includegraphics[width=0.4\textwidth]{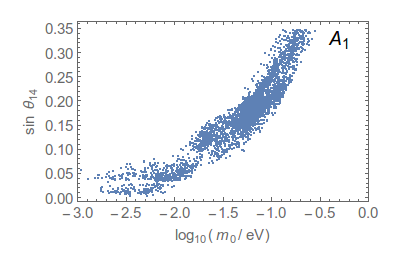}
\includegraphics[width=0.4\textwidth]{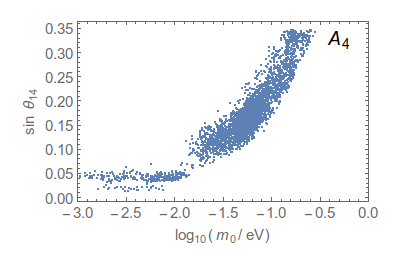} \\
\includegraphics[width=0.4\textwidth]{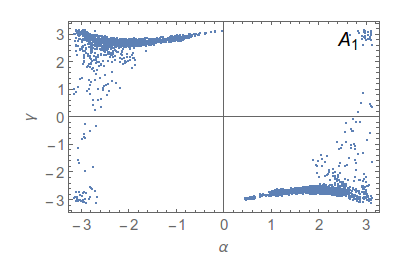}
\includegraphics[width=0.4\textwidth]{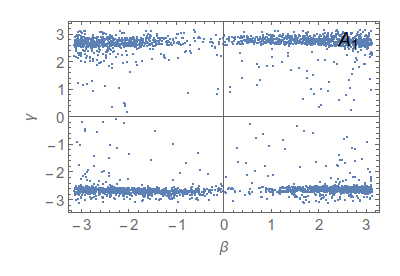}
\end{center}
\begin{center}
\caption{Correlation plots in class $A$ for NH. Top right panel is for texture $A_4$ and others are for texture $A_1$.}
\label{A_N}
\end{center}
\end{figure*}

In class $A$ with normal hierarchy, the allowed textures are $A_1$, $A_2$, $A_4$, $A_5$, $A_6$, $A_7$, $A_8$, $A_9$ and $A_{10}$. 
As all these textures have the condition $M_{ee}=0$,
%it will be sufficient to 
%\red{(Is it sufficient? Some correlations which we are not discussing might come about from the other zero elements, right? 
%Should we just say ``we analyze''?)} 
we analyze the mass matrix element $M_{ee}$ to understand the 
viability of these textures and correlation between different parameters. The expression for $M_{ee}$ according
to our parametrization is:
\begin{eqnarray}
\label{m}
 M_{ee} &=& c_{12}^2 c_{13}^2 c_{14}^2 m_1+e^{- i \alpha } c_{13}^2 c_{14}^2 m_2 s_{12}^2 \\ \nonumber
 &+& e^{- i \beta } c_{14}^2 m_3 s_{13}^2+e^{- i \gamma } m_4 s_{14}^2.
\end{eqnarray}
The sterile term in the above equation is given by  $m_4 s_{14}^2$. Note that though $m_4$ is large as compared to $m_i$ (i = 1,2,3), but the sterile term can be small if $s_{14}^2$ is not very large. 
Putting $m_1 = 0$ and $c_{13}=c_{14}=1$ (since $\theta_{13}$ and $\theta_{14}$ are small) in the above equation we obtain
\begin{eqnarray}
 M_{ee} = e^{- i \alpha } m_2 s_{12}^2 + e^{- i \beta } m_3 s_{13}^2+e^{- i \gamma } m_4 s_{14}^2.
 \label{m_n}
\end{eqnarray}
From this equation it is clear that when the lowest neutrino mass vanishes, one needs very small values of $\theta_{14}$ for cancellation to
occur between the active and sterile terms. 
%For non-zero $m_1$, the
%magnitude of the active term will become large and as
 For non-zero small values of $m_1$, the active term can still be small depending upon the values of the Majorana phases
and thus one can have cancellations for small values of $\theta_{14}$.
Now for large values of $m_1$, the active term will be dominated by the $m_1$ term and thus as $m_1$ increases,
one needs higher values of $\theta_{14}$ to obtain cancellation between 
the active and sterile terms. So we expect that in the allowed textures of
 class $A$, $\theta_{14}$ will be flat for small values of the lowest mass and then it will be an increasing function.
 This is clearly reflected in the 
upper panels of Fig.~\ref{A_N}. The left panel is for texture $A_1$ and
 the right panel is for texture $A_4$. 
 In both the panels we see that
  $\sin\theta_{14}$ remains almost constant for $-3 < \log_{10}(m_0) < -2$ and then it rises as the lowest mass $m_0$ ($= m_1$ for NH) increases.
Note that we have followed a conservative approach for lower bounds on
 active sterile mixing angles by taking them to be zero. However if one
  takes a lower allowed value of $\theta_{14}$ that is greater than zero
   from the global analysis~\cite{globalfit}, then we will not get $M_{ee} = 0$ for very small values of
$m_1$ and NH will be excluded in this region where $m_1 \rightarrow 0$.

%The same correlation is obtained for all the allowed texture in $A$ class for normal hierarcgy. 
From Eq.~\ref{m_n}, one can also understand that, in these cases, the Majorana phases have an
important role to play in obtaining $M_{ee}=0$. In the lower panels of Fig.~\ref{A_N}, we have plotted the 
Majorana phase $\gamma$ for the $A_1$ texture. From these plots we can see that the value
of $\gamma$ is constrained around $\pm \pi$ whereas the phases $\alpha$ and $\beta$ are unconstrained. 
This is expected because for these values of $\gamma$, 
the sterile term acquires a negative sign which is required for cancellation with the active terms to achieve $M_{ee}=0$. These correlations are characteristic of all the allowed textures in class $A$ for normal hierarchy.

\begin{figure*}
\begin{center}
\includegraphics[width=0.4\textwidth]{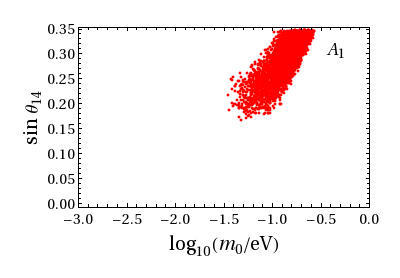}
\includegraphics[width=0.4\textwidth]{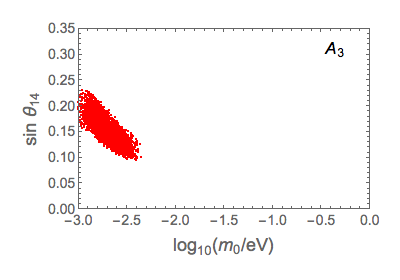} \\
\includegraphics[width=0.4\textwidth]{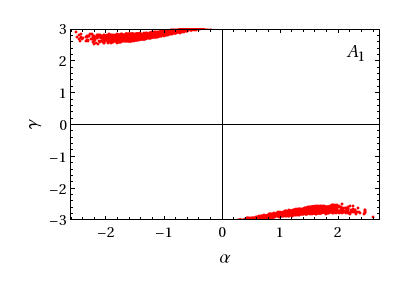}
\includegraphics[width=0.4\textwidth]{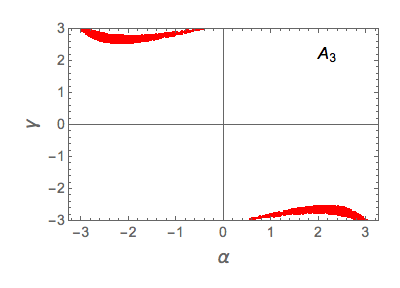}
\end{center}
\begin{center}
\caption{Correlation plots in class $A$ for IH. Left column corresponds to texture $A_1$ and the right column corresponds to texture $A_3$.}
\label{A_I}
\end{center}
\end{figure*}
\begin{figure*}
\begin{center}
\includegraphics[width=0.4\textwidth]{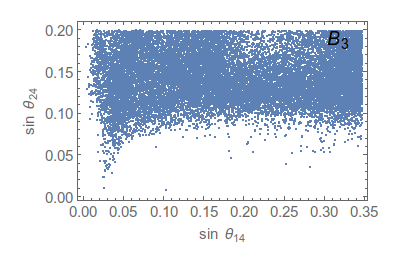}
\includegraphics[width=0.4\textwidth]{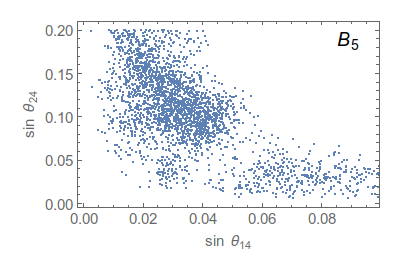} \\
\includegraphics[width=0.4\textwidth]{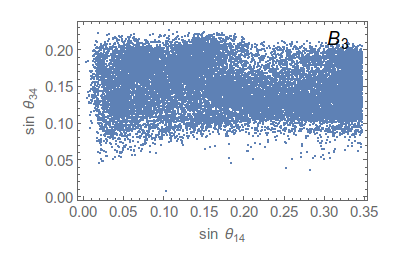}
\includegraphics[width=0.4\textwidth]{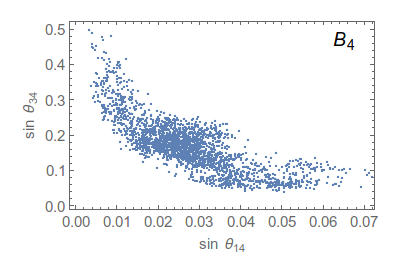}
\end{center}
\begin{center}
\caption{Correlation plots in class $B$ for NH. Left column corresponds to $B_3$ texture. The top right panel corresponds to $B_5$ texture and bottom right panel correspond to $B_4$ texture.}
\label{B_N}
\end{center}
\end{figure*}
 
 \begin{figure*}
\begin{center}
\includegraphics[width=0.4\textwidth]{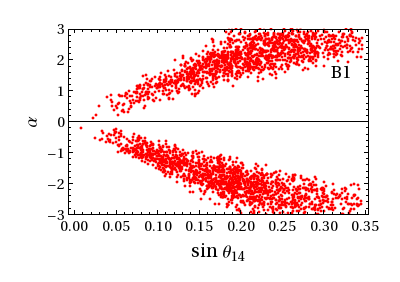}
\includegraphics[width=0.4\textwidth]{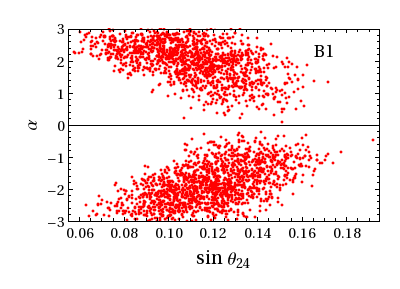} \\
\includegraphics[width=0.4\textwidth]{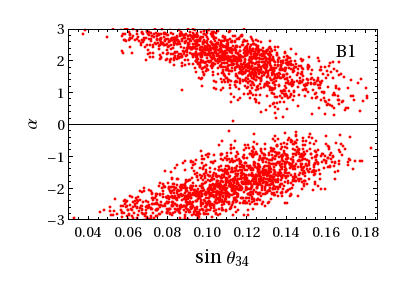}
\includegraphics[width=0.4\textwidth]{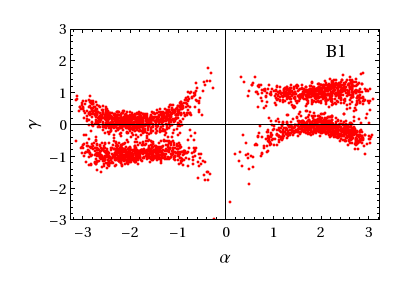}
\end{center}
\begin{center}
\caption{Correlation plots in class $B$ for IH. All plots are for texture $B_1$.}
\label{B_I}
\end{center}
\end{figure*}
The conclusions for inverted hierarchy in class $A$ are quite similar to normal hierarchy. For inverted hierarchy, the allowed textures in class $A$ are
$A_1$, $A_2$, $A_3$, $A_4$, $A_5$, $A_6$, $A_7$ and $A_8$. 
Among these, the texture $A_3$ is allowed only for small $m_3$ values whereas $A_6$ is allowed for both smaller and higher values of $m_3$. Apart from $A_3$ and $A_6$, we find that 
lower values of $m_3$ are not allowed in any of the remaining textures in $A$ class for IH.
%Putting $m_3=0$ in Eq.~\ref{m} with $c_{13}=c_{14}=1$ and 
%using Eq.~\ref{ih}  we obtain
% \begin{eqnarray}
%  \label{m_ii}
%  M_{ee} &=& c_{12}^2 \sqrt{|\Delta m_{32}^2|} \\ \nonumber
%  &+& e^{- i \alpha } \sqrt{|\Delta m_{32}^2|} s_{12}^2 + e^{- i \gamma } m_4 s_{14}^2.
% \end{eqnarray}
%Here for smaller values of m_3, one can have M_{ee}=0 for IH
%In the above equation we used the fact that $\Delta m_{21}^2 \ll \Delta m_{32}^2$ which also implies $m_1 = m_2$.
%Here when $m_3=0$, the active terms are quite large due to the dominant effect of the atmospheric mass squared difference. 
%Thus very small values of $\theta_{14}$ can not produce
%$M_{ee}=0$. But similar to NH, as $m_3$ increases from zero, we need higher values of $\theta_{14}$ for cancellation 
%to occur. 
%This can be seen from the upper panels of
In Fig. \ref{A_I}, we have plotted the correlations between the different parameters for IH. The left panel is
for texture $A_1$ and the right panel is for texture $A_3$. 
From the upper left panel we can see that $\sin\theta_{14}$ 
is a rising function of the lowest mass $m_0$ ($=m_3$ for IH) and in the upper right panel an anti-correlation between $\sin\theta_{14}$ and $m_0$ is observed.
As mentioned earlier, from the figures we also see that for the texture $A_1$ lower values of $m_0$ are not allowed and the texture $A_3$ is allowed only for lower values of $m_0$.
%values of $m_0$ are not allowed for majority of textures
%of class $A$ except textures $A_3$ and $A_6$. 
%Note that though the element $M_{ee}$ can be zero
%for smaller values of $m_3$ \cite{3+11zero} for IH, but in our case due to the interplay of the other elements, the lower values of $m_3$ are getting disallowed for textures $A_1$,
%$A_2$, $A_4$, $A_5$, $A_7$ and $A_8$. 
%For textures $A_3$ and $A_6$, however, very small values of $m_3$ 
%are allowed as can be seen from upper right panel of Fig. \ref{A_I}. 
%In both the plots we can see that $\sin\theta_{14}$ 
%is a rising function of the lowest mass $m_0$ ($=m_3$ for IH) but unlike normal hierarchy, very small
%values of $\sin\theta_{14}$ are not allowed. 
%Another important point to note is that for IH, 
%the expression for $M_{ee}$ (cf. Eq.~\ref{m_ii}) contains only Majorana phases $\alpha$ and $\gamma$. Thus in this case
%we expect to obtain a strong correlation between these two phases.
Similar to NH, here also we found that the Majorana phase $\gamma$ is strongly constrained around $\pm \pi$ (lower panels of Fig. \ref{A_I}).
%This
% is evident from the lower panels of Fig. \ref{A_I}. Here we see that for
%  both the textures i.e., $A_1$ and $A_4$,
%$\gamma$ is strictly restricted in $\pm \pi$.
Though we have shown our results only for texture $A_1$ and $A_4$, the conclusions drawn in this section 
are also applicable for all the other allowed textures in class $A$.

\subsection{Disallowed textures in class $A$}

For NH in class $A$ , only $A_3$ texture is phenomenologically disallowed. 
Apart from $M_{ee}=0$, this texture satisfies the condition 
$M_{\mu \mu} = M_{\mu \tau} = M_{\tau \tau} = 0$. To explain
 why this texture is disallowed for NH but allowed IH, we refer to the
lower panels of Fig. 5 of Ref.~\cite{3+13zero}.
In those panels, it is shown that $M_{\mu \mu} = M_{\mu \tau} = M_{\tau \tau} = 0$ 
($=C_3$, following the notation of this reference) is only possible for NH 
if $\theta_{14}$ has very large values $(\theta_{14} > 50^\circ)$. Such large values 
are disallowed by the current data. 
But for IH, the allowed range for $\theta_{14}$ with $C_3=0$, is 
$0^\circ < \theta_{14} < 80^\circ$ which includes values of $\theta_{14}$
allowed by the current data.
This explains why the texture $A_3$ is disallowed for NH but allowed for IH.

For IH, the disallowed textures in class $A$ are $A_9$ and $A_{10}$. 
Apart from $M_{ee}=0$, in $A_9$ we have the condition $E_3 = M_{e \tau} = M_{\mu \tau} = M_{\tau \tau}=0$ 
and in $A_{10}$, we have 
$F=M_{e \mu} = M_{e \tau} = M_{\mu \tau} = 0$. The results of Ref.~\cite{3+13zero} 
show that both $E_3$ = 0 and $F=0$ are allowed for NH and IH. But $E_3 = 0$ for IH and $F = 0$, prefer
$\alpha= \pi$ (cf. Table VI of Ref.~\cite{3+13zero}). From the lower panels 
of Fig.~\ref{A_I}, we note that $\alpha=0$ and $\alpha=\pi$ are almost disallowed 
for IH in class $A$. This explains why the textures
$A_9$ and $A_{10}$ are not allowed for IH.

\subsection{Allowed textures in class $B$}

\begin{figure*}
\begin{center}
\includegraphics[width=0.4\textwidth]{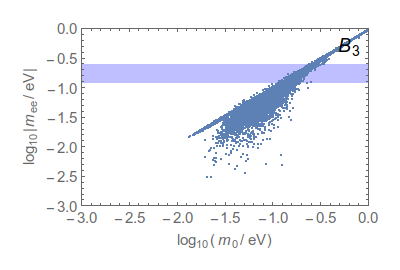}
\includegraphics[width=0.4\textwidth]{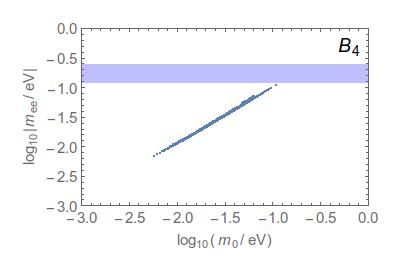} \\
\includegraphics[width=0.4\textwidth]{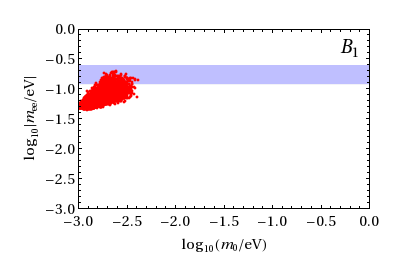}
\includegraphics[width=0.4\textwidth]{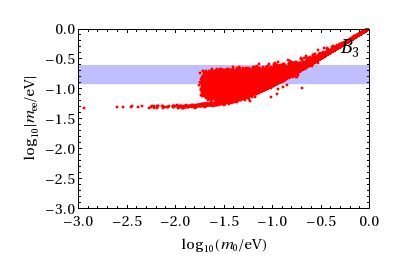} \\
\includegraphics[width=0.4\textwidth]{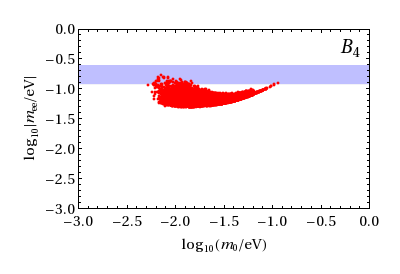}
\includegraphics[width=0.4\textwidth]{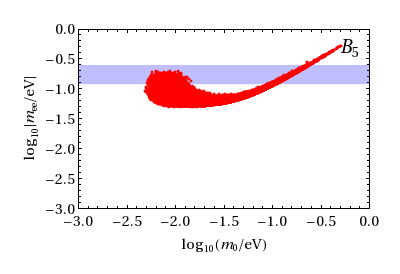}\\
\end{center}
\begin{center}
\caption{$M_{ee}$ prediction plots in class $B$. The upper panels are for NH. 
The middle and lower panels are for IH. The blue horizontal band corresponds to the current experimental bound on $M_{ee}$ from combined
results of KamLAND and EXO-200~\cite{kamland_zen}.
}
\label{mee}
\end{center}
\end{figure*}
In class $B$ for normal hierarchy, the allowed textures are $B_3$, $B_4$ and $B_5$. 
In all these three textures we have $M_{e\mu}=M_{e\tau}=0$. 
From the expressions of $M_{e \mu}$ and $M_{e \tau}$ (cf. Appendix~\ref{appen1}), 
we can see that in $M_{e \mu}$, the $m_4$ term appears with $s_{14} s_{24}$ and 
in $M_{e \tau}$, $m_4$ appears with
 $s_{14} s_{34}$. 
%  {\bf Note that, the elements $M_{e\mu}$ and $M_{e \tau}$ (also $M_{\mu \mu}$ and $M_{\tau \tau}$) are related 
%  by $\mu-\tau$ exchange symmetry given by \cite{3+11zero}:
%  \begin{equation}
% \bar\theta_{12}= \theta_{12},
% ~~~ \bar\theta_{13} = \theta_{13},
% ~~~ \bar\theta_{14} = \theta_{14},
% \end{equation}
% \begin{equation} \label{mutau1}
% \sin{\bar\theta_{24}} =  \sin\theta_{34} \cos{\theta_{24}},
% \end{equation}
% \begin{equation}
% \sin {\bar\theta_{23}}
% =\frac{\cos{\theta_{23}}\cos{\theta_{34}}-\sin{\theta_{23}}\sin{\theta_{34}}\sin{\theta_{24}}}{\sqrt{1-\cos{\theta_{24}^2}\sin{\theta_{34}^2}}},
% \end{equation}
% \begin{equation}
% \sin {\bar\theta_{34}} \label{mutau2}
% =\frac{\sin{\theta_{24}}}{\sqrt{1-\cos{\theta_{24}^2}\sin{\theta_{34}^2}}}.
% \end{equation}
% It is easy to show that for small values of $\theta_{24}$ and $\theta_{34}$:
% \begin{eqnarray}
% \bar\theta_{24} \approx \theta_{34} \\
% \bar\theta_{34} \approx \theta_{24}
% \end{eqnarray}
% Thus we expect that the behaviour of
%  $\theta_{24}$ in $M_{e \mu}$ will be similar as that of behaviour of 
%  $\theta_{34}$ in $M_{e \tau}$~\cite{3+11zero}. }
 Thus we understand that to achieve cancellation between the active terms and the sterile term,  
 either $s_{14}$ or one of $s_{24}$ and $s_{34}$
 has to always be small for completely hierarchical mass spectrum.
 In Fig.~\ref{B_N}, we have given the correlations plots for the class $B$. 
 The left column corresponds to the $B_3$ texture.
 In both the panels we can see that $\sin\theta_{24}$ and $\sin\theta_{34}$ prefer 
 to have values on the higher side, whereas $\sin\theta_{14}$ remains unconstrained. In our analysis
 we find that the nature of correlation between $\sin\theta_{24}$ ($\sin\theta_{34}$) 
 and $\sin\theta_{14}$ for $B_3$ texture is similar as that of $B_4$ ($B_5$) texture. 
 But the correlation between $\sin\theta_{24}$ ($\sin\theta_{34}$) and $\sin\theta_{14}$ 
 in $B_5$ ($B_4$) is different as can be seen from the top right (bottom right) panel of Fig. \ref{B_N}.
 From these figures we can see that, $\sin\theta_{24}$ ($\sin\theta_{34}$) and 
 $\sin\theta_{14}$ can not both have very large or very small values 
 simultaneously in $B_5$ ($B_4$) textures.

For IH, all the textures in class $B$ are allowed. The texture $B_1$ contains 
$M_{e \mu}=0$ and $B_2$ contains $M_{e \tau}=0$, whereas $B_3$, $B_4$ and $B_5$ contain $M_{e \mu}=M_{e \tau} =0$.
The full expression for $M_{e \mu}$ according to our parametrization is given by
\begin{widetext}
\begin{eqnarray}
 M_{e\mu} &=&-e^{-i \delta _{24}} c_{14} \big(e^{i \delta _{24}} c_{12} c_{13} c_{23} c_{24} \big(m_1-e^{- i \alpha } m_2\big) s_{12} - e^{i \big(\delta_{13}+\delta _{24}\big)}
 c_{13} c_{24} \big(e^{- i \beta } m_3-e^{- i \alpha } m_2 s_{12}^2\big) s_{13} s_{23} \\ \nonumber 
 &+& e^{i \big(2 \alpha +\delta _{14}\big)} c_{13}^2 m_2 s_{12}^2 s_{14} s_{24}-e^{i \delta _{14}} \big(e^{- i \gamma } m_4 - e^{- i \beta } m_3 s_{13}^2\big) s_{14} s_{24}+c_{12}^2 c_{13}
m_1 \big(e^{i \big(\delta _{13}+\delta _{24}\big)} c_{24} s_{13} s_{23}  +e^{i \delta _{14}} c_{13} s_{14} s_{24}\big)\big)
\end{eqnarray}
Putting $m_3=0$  with $m_1=m_2 =\sqrt{|\Delta m_{32}^2|}$ (using $\Delta m_{21}^2 \ll \Delta m_{32}^2$), $c_{13}=c_{14}=c_{24}=1$ and keeping terms up to the second order
of the small parameters $\theta_{14}$, $\theta_{24}$ and $\theta_{13}$, $M_{e \mu}=0$ condition reduces to
\begin{equation}  \label{memih}
 M_{e \mu} = c_{12}s_{12}c_{23}(e^{-i\alpha}-1)- s_{23}s_{13}e^{i\delta_{13}}(c_{12}^2+s_{12}^2 e^{-i\alpha}) - e^{i(\delta_{14}-\delta_{24})} s_{14}s_{24}
 \bigg(c_{12}^2 -e^{-i\gamma}\sqrt{\frac{\Delta m_{43}^2}{|\Delta m_{32}^2|}} + e^{-i\alpha}s_{12}^2\bigg)=0
\end{equation}
\end{widetext}
Now further putting $\alpha = \gamma = 0$ and all the Dirac phases as $\pi$, we obtain
\bea
 s_{23}s_{13}-(1-\sqrt{\frac{\Delta m_{43}^2}{|\Delta m_{32}^2|}})s_{14}s_{24}=0.
\eea
Under the similar approximation, the simplified expression of $M_{e \tau}=0$ can be written as
\bea
&& s_{13} (c_{23}c_{34} +s_{23}s_{34}s_{24}) + \\ \nonumber
&& s_{34}s_{14}(1-\sqrt{\frac{\Delta m_{43}^2}{|\Delta m_{32}^2|}}) = 0.
\eea
%The similarity of both these equations are due to $\mu$-$\tau$ symmetry. 
From both the equations we notice that, when the phase $\alpha$ goes to $0$, 
there are no leading order terms i.e.,
terms that contain the small parameters $s_{13}$, $s_{24}$ and $s_{14}$ are absent. 
Thus for inverted hierarchy, to obtain four-zero texture in class $B$ one of the angles that appears with the
$m_4$ term in $M_{e \mu}$ and $M_{e \tau}$ has to be very small at $\alpha=0$. In Fig.~\ref{B_I}, 
we present the correlation plots for $B_1$ texture. From these plots we see 
that when the phase $\alpha$ is equals to zero,
$\sin\theta_{14}$ is very small but as $\alpha$ deviates from zero, larger 
values of $\sin\theta_{14}$ get preferred (top left panel of Fig.~\ref{B_I}).
But on the other hand, we observe the reverse features for $\sin\theta_{24}$ 
and $\sin\theta_{34}$. For $\alpha=0$, these angles prefer 
higher values and for $\alpha = \pm \pi$, lower values of
$\sin\theta_{14}$ and $\sin\theta_{24}$ get allowed (top right and bottom 
left panels of Fig.~\ref{B_I}). Note that when $m_3 = 0$, the equations 
of $M_{e \mu}$ and $M_{e \tau}$ contains the Majorana phases
$\alpha$ and $\gamma$. Thus to see their nature, in the bottom right panel of 
Fig.~\ref{B_I}, we present the correlation plot in the $\alpha$ -$\gamma$ plane for the $B_1$ texture. 
In that plot we see that $\gamma = \pm \pi$ is strictly disallowed.
The correlations plots
presented in Fig.~\ref{B_I} are similar for all the allowed textures in $B$ class for inverted hierarchy.

\subsection{Disallowed textures in class $B$}

In class $B$, the textures $B_1$ and $B_2$ are not allowed for NH but they are allowed for IH. 
This is simply because, these textures also contains the structure $C_3 = 0$ which is disallowed in NH and allowed in
IH for the values of $\theta_{14}$ considered in our analysis. For IH, there are no disallowed textures in class $B$. 

\subsection{Implications for neutrinoless double beta decay} 

The effective Majorana mass $M_{ee}$ is important because
the neutrinoless double beta decay (NDBD) experiments can give an upper 
bound on $M_{ee}$. The latest bound on $M_{ee}$ comes from the combined 
analysis of KamLAND-ZEN and
EXO-200~\cite{kamland_zen} and the value turns out to be $M_{ee}<$
 (0.12 - 0.25) eV at $90\%$ C.L, 
 where the width arises due to the uncertainty in the
 nuclear matrix elements\footnote{According to the most recent KamLAND-ZEN results, 
 the upper bound on $M_{ee}$ is (0.06 - 0.16) eV at $90\%$ C.L.~\cite{KamLAND-Zen:2016pfg}.}. 
 %\red{(KamLAND-ZEN has recently given out better results: 1605.02889 . 
 %We should mention it, maybe put an extra line/band in the plots of Fig 5 
 %and see if some physics changes. For example I think a very small 
 %part of the B1 and B3 parameter space (middle row) will be allowed.)}
Finally in Fig.~\ref{mee}, we present the predictions for the effective 
Majorana mass $M_{ee}$ for all the allowed textures in $B$ class. 
For NH, we have given the plots for
the textures $B_3$ and $B_4$ (upper panels). 
The prediction for $B_5$ is similar as that of $B_4$. For IH, we present the
same for the textures $B_1, B_3, B_4, B_5$. We do not show the
corresponding plot for the texture $B_2$ as this has the same
prediction for $M_{ee}$ as that of texture $B_1$. For $B_4$ (and $B_5$)
with NH, it is seen from top right panel of Fig.~\ref{mee} that the
 predictions for $M_{ee}$ lie below the current experimental bound for
 lightest neutrino mass up to around $0.1$ eV. On the other hand, for
 $B_3$ texture with NH, certain predictions for $M_{ee}$ will be ruled
 out by current bounds. These values of $M_{ee}$ correspond to higher
 values of lightest neutrino mass $m_0$ (in this case $m_1$) $\geq 0.1$ eV. In the case of IH,
 the predicted values of $M_{ee}$ are either very close to the
 present experimental bound or ruled out by current
bounds for larger values of the lightest mass (in this case $m_3$). In all these textures there 
 are values of $M_{ee}$ that lie below the present experimental
 bound when the lightest mass is small.
 For textures $B_1, B_4, B_5$ with IH, certain values of the lightest
 neutrino mass smaller than $0.01$ eV can get ruled out from the
 current experimental bounds, as can be seen from Fig.~\ref{mee}.

\section{Flavor symmetry origin of four-zero texture}
\label{symmetry}

\begin{table}
\begin{center}
\begin{tabular}{|c|c|c|}
\hline
Fermion Fields & $SU(3)_c \times SU(2)_L \times U(1)_Y$ & $Z_3 \times Z_3$ \\
\hline
$ L_e$ & $ (1, 2, -1)$ & $(\omega, 1)$ \\
$ L_\mu$ & $ (1, 2, -1)$ & $(\omega, 1)$ \\
$ L_\tau$ & $ (1, 2, -1)$ & $(\omega^2, 1)$ \\
$ \nu_s $ & $ (1, 1, 0)$ & $(1, \omega)$ \\
\hline
Scalar Fields & $SU(3)_c \times SU(2)_L \times U(1)_Y$ & $Z_3 \times Z_3$ \\
\hline
$ \Delta_1 $ & $ (1, 3, 2)$ & $(1, 1)$ \\
$ \Delta_2$ & $ (1, 3, 2)$ & $(\omega^2, \omega)$ \\
$ \Delta_3$ & $ (1, 3, 2)$ & $(\omega, \omega)$ \\
$ S_1 $ & $ (1, 2, -1)$ & $(1, \omega)$ \\
$ H $ & $(1,2,1)$ & $(1,1)$ \\
$ \chi $ & $(1,1,0)$ & $(\omega, 1)$\\
\hline
\end{tabular}
\end{center}
\caption{Fields responsible for $4\times4$ light neutrino mass matrix with four-zero texture}
\label{tab3}
\end{table}
Since the zero textures appear only in the $3\times3$ block of  the $4\times4$ mass matrix, they can be explained by 
different flavor symmetry frameworks. Some possible models are discussed in 
Ref.~\cite{texturesym,texturesym1,texturesym2,texturesym3,texturesym4,texturesym5,texturesym6,texturesym7,texturesym8,texturesym9} in 
the context of zero textures in three neutrino picture. 
As an illustrative example, here we show how a particular four-zero texture can be realized within 
type II seesaw framework~\cite{tii,tii1,tii2,tii3,tii4,tii5,tii6}. The mixing terms between active and sterile neutrino sectors must however, 
arise from a different mechanism as we can not have a type II seesaw term for active-sterile or sterile-sterile mass terms due to 
the singlet nature of sterile neutrinos under the electroweak gauge symmetry. 
Several new physics mechanisms have been proposed
recently~\cite{sterileearlier1, sterileearlier2, sterileearlier3, sterileearlier4, sterileearlier5, sterileearlier6, sterileearlier7, sterileearlier8, sterileearlier9, sterileearlier10} 
in order to accommodate light sterile neutrinos with masses in the eV-keV range. 
In a model with type II seesaw for active neutrino masses, if we add a sterile neutrino, 
one can in principle have tree level active-sterile mixing through the standard model Higgs and 
also a bare mass term for the sterile neutrino. However, to keep the sterile neutrino mass at eV scale and 
active-sterile mixing terms at sub-eV scale, one needs to tune the Yukawa couplings and the bare mass term in an unnatural way.
This invokes the presence of additional new dynamics responsible for such tiny mass matrix elements. 
In this section, we show the active-sterile and sterile-sterile interaction terms to originate 
from an effective theory point of view without choosing a particular UV complete theory of new physics. 
A discrete flavor symmetry and the transformation of the fields under this symmetry are chosen in such way that it 
generates the desired texture zeros in the active neutrino block through tree level type II seesaw, and active-sterile, 
sterile-sterile mass terms arise only through dimension five operators.

The four-zero texture pattern in the mass matrix denoted by $A_1$ above can be explained 
by considering an additional discrete symmetry $Z_3 \times Z_3$ and a few scalar fields 
whose transformations are shown in table~\ref{tab3}. This allows us to write the following 
Yukawa Lagrangian up to dimension five responsible for the desired four-zero texture of the mass matrix $A_1$.
\begin{widetext}
\begin{align}
\mathcal{L}_{\text{Yukawa}} & \subset Y_{e\tau} L^T_e C i \sigma_2\Delta_1 L_{\tau} + Y_{\mu\tau} L^T_\mu C i \sigma_2\Delta_1 L_{\tau} + \frac{Y_{es}}{\Lambda} L^T_e C i \sigma_2\Delta_2 S_1 \nu_s+\frac{Y_{\mu s}}{\Lambda} L^T_\mu C i \sigma_2\Delta_2 S_1 \nu_s \nonumber \\
& + \frac{Y_{\tau s}}{\Lambda} L^T_{\tau} C i \sigma_2\Delta_3 S_1 \nu_s + \frac{Y_{ss}}{\Lambda} \nu_s \nu_s S_1 H
\end{align}
\end{widetext}
The relevant part of the scalar Lagrangian can be written as
\begin{widetext}
\begin{align}
\mathcal{L}_{\text{Scalar}} & \subset -\mu^2_H H^{\dagger} H + \frac{\lambda_H}{4} (H^{\dagger} H)^2 -\mu^2_S S^{\dagger}_1 S_1 + \frac{\lambda_S}{4} (S^{\dagger}_1 S_1)^2 +\mu^2_{\Delta_i} \Delta^{\dagger}_i \Delta_i + \frac{\lambda_{\Delta_i}}{4} (\Delta^{\dagger}_i \Delta_i)^2-\mu^2_{\chi} \chi^{\dagger} \chi + \frac{\lambda_{\chi}}{4} (\chi^{\dagger} \chi)^2 \nonumber \\
& + (\mu_1 H^T \Delta^{\dagger}_1 H + \lambda_1 S^T_1 \Delta_2 S_1 \chi + \lambda_2 S^T_1 \Delta_3 S_1 \frac{\chi^2}{\Lambda} + \text{h.c.})
\end{align}
\end{widetext}
The neutral components of scalar doublets $H, S_1$ acquires non-zero vacuum expectation value (vev) leading to spontaneous electroweak symmetry breaking and non-zero charged fermion masses. 
The neutrino masses are however, proportional to the vev's of the neutral components of the scalar triplet fields. Precision constraints on the electroweak $\rho$ parameter restricts the triplet vev to $v_L \leq 2$ GeV \cite{rhoparam}. 
This can be satisfied naturally in the model, if the neutral component of the scalar triplet acquires an induced vev after the electroweak symmetry breaking. 
Minimization of the above scalar Lagrangian with respect to these neutral components of the triplets give rise to 
\begin{widetext}
\begin{equation}
 \langle \Delta^0_1 \rangle \approx \frac{\mu_1 \langle H^0 \rangle^2}{\mu^2_{\Delta_1}}, \;\; \langle \Delta^0_2 \rangle \approx \frac{\lambda_1 \langle S^0_1 \rangle^2 \langle \chi \rangle}{\mu^2_{\Delta_2}}, \;\; \langle \Delta^0_3 \rangle \approx \frac{\lambda_2 \langle S^0_1 \rangle^2 \langle \chi \rangle^2}{\mu^2_{\Delta_3} \Lambda}
 \label{veveq}
 \end{equation}
\end{widetext}
It can be seen that the neutral component of the scalar triplet $\Delta_1$ can acquire an induced vev after the scalar doublet $H$ acquires a non-zero vev leading to electroweak symmetry breaking. 
This can generate the first $3\times 3$ block of the $4\times 4$ light neutrino mass matrix. 
However, in order to generate the active-sterile and sterile-sterile mass terms, the neutral components of the other two scalar triplets $\Delta_{2,3}$ must acquire a tiny induced vev. 
As it turns out, this is possible only when both the scalar doublet $S_1$ and the scalar singlet $\chi$ acquire non-zero vev's which thereby induce a tiny vev to the neutral components of $\Delta_{2,3}$ as can be seen from equation \eqref{veveq}.
The trilinear and quartic couplings as well as the bare mass terms of the triplet scalars in the above expressions \eqref{veveq} can be adjusted 
in such a way that results in tiny triplet vev's, generating the elements of the $4\times 4$ light neutrino mass matrix. The first $3\times3$ block of mass matrix $A_1$ is generated by ordinary type II seesaw mechanism whereas 
the terms involving sterile neutrino $\nu_s$ arise from dimension five effective terms indicating the presence of 
new physics at cut-off scale $\Lambda$ responsible for tiny sterile neutrino mass and its mixing with active neutrinos. 
 The additional discrete symmetry $Z_3 \times Z_3$ is chosen such that it does not allow the bare mass term of 
sterile neutrino $M_s \nu_s \nu_s$, as it would be unnatural to have $M_s$ at eV scale in that case.
The vacuum expectation values of three triplet and two doublet scalars can be adjusted in a way that the above terms 
give rise to a $4\times4$ Majorana neutrino mass matrix with all entries at sub-eV to eV scale apart from the texture zero elements. 
A full discussion on UV completeness of such scenarios can be found elsewhere and we do not pursue it further in this work.

\section{Summary}
\label{sum}
In this work, we have studied the possibility of having four zeros in the low energy neutrino mass matrix where the light neutrino sector consists 
of three active and one sterile neutrino at eV scale. Considering the neutrinos to be Majorana particles resulting in a complex 
symmetric $4\times 4$ mass matrix, we parametrise it using four mass eigenvalues, six mixing angles and six phases: a total of 
sixteen parameters. Using the global fit $3\sigma$ values of eight parameters namely, three mass squared differences, 
three active neutrino mixing angles and two active-sterile mixing angles, we solve for the other eight parameters: lightest neutrino mass, 
six phases and one active-sterile mixing angle by using the eight real constraint equations from four texture zeros. 
Using the earlier results which disfavoured zero entries in $M_{\alpha s}$, where $\alpha=e,\mu,\tau$, we consider the fifteen possible 
four-zero textures in the $3\times3$ active neutrino block of the $4\times4$ mass matrix. 
We find that, out of these fifteen possibilities, only twelve are allowed in 
NH whereas thirteen textures are allowed in IH. We have summarized our results in table~\ref{su}. 
Apart from studying the viability of the possible textures, we also find the predictions for specific 
neutrino parameters as well as effective neutrino mass for neutrinoless double beta decay. 
Some of these correlations are shown in Figs.~\ref{A_N},~\ref{A_I},~\ref{B_N},~\ref{B_I} and~\ref{mee}. 
We explain the correlations from the analytical expressions of the relevant mass matrix elements. 
In class $A$, one can easily understand the 
correlations shown in Figs.~\ref{A_N},~\ref{A_I}
just by analyzing the element $M_{ee}$.  
Similarly, we present the analytical understanding of the correlation plots shown for the class $B$ textures (shown in Figs.~\ref{B_N},~\ref{B_I})
by analyzing the mass matrix elements
$M_{e \mu}$ and $M_{e \tau}$. 
%We point out the $\mu-\tau$ exchange symmetry between $M_{e\mu}=0$ and $M_{e\tau}=0$ which explains the similar 
%behavior of $\theta_{24}$ in $M_{e\mu}$ and $\theta_{34}$ in $M_{e\tau}$ for NH, as seen from Fig.~\ref{B_N}. 
%The same exchange symmetry is also pointed out for IH cases of class $B$ textures in order to understand the correlation plots shown in Fig.~\ref{B_I}.
While studying the prediction for the effective Majorana mass $M_{ee}$ in $B$ class, we found that 
some regions of the allowed parameter space fall in the range of current experimental sensitivity of neutrinoless double beta decay apart 
from ruling out the regions for high values of lightest neutrino mass $m_0$ ($m_1$ for NH and $m_3$ for IH) $ \geq 0.1$ eV. 
In the end, we briefly outline a possible way of generating one such four-zero texture mass matrix using an effective theory framework with some discrete flavor symmetry. 
The $3\times3$ active neutrino block with four zeros is shown to arise from a tree level 
type II seesaw mechanism whereas the active-sterile block is shown to arise from 
dimension-five effective terms. We do not discuss the details of the UV complete theory responsible for eV scale sterile neutrino masses and leave it for future investigations.

In summary we say that the present status of light sterile neutrinos is still very intriguing and future experiments are expected to shed more light into these scenarios. On the other hand, 
future observation or non-observation of neutrinoless double beta decay along with the measurements of neutrino mass hierarchy, 
can also give an insight towards the possibility of the light sterile neutrinos and hence can probe the viability of the texture zeros in the low energy neutrino mass matrix. 
The results discussed in this work should be able to guide future model building works on eV scale sterile neutrino masses and their mixing with active neutrinos.

\section*{Acknowledgements}
The authors would like to thank Srubabati Goswami for suggesting this problem in WHEPP XIII at Puri in December 2013.
The authors MG and SG would like to thank Sanjib Kumar Agarwalla and Efunwande Osoba for useful discussions in WHEPP XIII.
DB, MG, SP and SKR would like to thank the organisers of WHEPP XIV for hospitality at IIT Kanpur in December, 2015 where this work is initiated.
MG would like to thank Abhay Swain for help in Mathematica.
The work of MG is partly supported by the ``Grant-in-Aid for Scientific Research of the Ministry of Education, 
Science and Culture, Japan", under Grant No. 25105009. The work of SG is supported by the Australian Research 
Council through the ARC Center of Excellence in Particle Physics (CoEPP Adelaide) at the Terascale (CE110001004).

\appendix
\section{Light neutrino mass matrix elements}
\label{appen1}

\begin{widetext}
\begin{equation}
M_{ee} = c_{12}^2 c_{13}^2 c_{14}^2 m_1+e^{- i \alpha } c_{13}^2 c_{14}^2 m_2 s_{12}^2+e^{- i \beta } c_{14}^2 m_3 s_{13}^2+e^{-i \gamma } m_4 s_{14}^2
\end{equation}
\begin{eqnarray}
M_{e\mu} &=&-e^{-i \delta _{24}} c_{14} \big(e^{i \delta _{24}} c_{12} c_{13} c_{23} c_{24} \big(m_1-e^{- i \alpha } m_2\big) s_{12}-e^{i \big(\delta
_{13}+\delta _{24}\big)} c_{13} c_{24} \big(e^{- i \beta } m_3-e^{- i \alpha } m_2 s_{12}^2\big) s_{13} s_{23} \nonumber \\
 && +e^{i \big(2 \alpha +\delta _{14}\big)}M
c_{13}^2 m_2 s_{12}^2 s_{14} s_{24}-e^{i \delta _{14}} \big(e^{- i \gamma } m_4-e^{- i \beta } m_3 s_{13}^2\big) s_{14} s_{24}+c_{12}^2 c_{13}
m_1 \big(e^{i \big(\delta _{13}+\delta _{24}\big)} c_{24} s_{13} s_{23} \nonumber  \\
&& +e^{i \delta _{14}} c_{13} s_{14} s_{24}\big)\big)
\end{eqnarray}
\begin{eqnarray}
M_{e\tau}&=&c_{14} \big(-e^{i \big(- \alpha +\delta _{14}\big)} c_{13}^2 c_{24} m_2 s_{12}^2 s_{14} s_{34}+e^{i \delta _{14}} c_{24} \big(e^{- i \gamma
} m_4-e^{- i \beta } m_3 s_{13}^2\big) s_{14} s_{34} \nonumber \\
&& +c_{12} c_{13} \big(m_1-e^{- i \alpha } m_2\big) s_{12} \big(c_{34} s_{23}+e^{i \delta
_{24}} c_{23} s_{24} s_{34}\big)+e^{i \delta _{13}} c_{13} \big(e^{- i \beta } m_3-e^{- i \alpha } m_2 s_{12}^2\big) s_{13} \big(c_{23} c_{34} \nonumber \\
&& -e^{i
\delta _{24}} s_{23} s_{24} s_{34}\big)-c_{12}^2 c_{13} m_1 \big(e^{i \delta _{13}} c_{23} c_{34} s_{13}+\big(e^{i \delta _{14}} c_{13} c_{24}
s_{14}-e^{i \big(\delta _{13} +\delta _{24}\big)} s_{13} s_{23} s_{24}\big) s_{34}\big)\big)
\end{eqnarray}
\begin{eqnarray}
M_{\mu\mu} &=&e^{ i \big(-\gamma + 2\delta_{14}- 2 \delta_{24}\big)} c_{14}^2 m_4 s_{24}^2+e^{- i \beta } m_3 \big(e^{i \delta _{13}} c_{13} c_{24} s_{23}-e^{i
\big(\delta _{14}-\delta _{24}\big)} s_{13} s_{14} s_{24}\big){}^2+e^{- i \alpha } m_2 \big(c_{12} c_{23} c_{24}\nonumber \\
&& +s_{12} \big(-e^{i \delta
_{13}} c_{24} s_{13} s_{23}-e^{i \big(\delta _{14}-\delta _{24}\big)} c_{13} s_{14} s_{24}\big)\big){}^2+m_1 \big(c_{23} c_{24} s_{12}+c_{12}
\big(e^{i \delta _{13}} c_{24} s_{13} s_{23} \nonumber \\
&& +e^{i \big(\delta _{14}-\delta _{24}\big)} c_{13} s_{14} s_{24}\big)\big){}^2
\end{eqnarray}
\begin{eqnarray}
M_{\mu\tau} &=& e^{i \big(- \gamma +2 \delta _{14}-\delta _{24}\big)} c_{14}^2 c_{24} m_4 s_{24} s_{34}+e^{i \big(2 \beta +\delta _{13}\big)} m_3 \big(e^{i
\delta _{13}} c_{13} c_{24} s_{23}-e^{i \big(\delta _{14}-\delta _{24}\big)} s_{13} s_{14} s_{24}\big)  \nonumber \\
&&\big(-e^{-i \big(\delta _{13}-\delta
_{14}\big)} c_{24} s_{13} s_{14} s_{34}+c_{13} \big(c_{23} c_{34}-e^{i \delta _{24}} s_{23} s_{24} s_{34}\big)\big)+m_1 \big(-c_{23} c_{24}
s_{12}+c_{12} \big(-e^{i \delta _{13}} c_{24} s_{13} s_{23} \nonumber \\
&& -e^{i \big(\delta _{14}-\delta _{24}\big)} c_{13} s_{14} s_{24}\big)\big) \big(s_{12}
\big(c_{34} s_{23}+e^{i \delta _{24}} c_{23} s_{24} s_{34}\big)+c_{12} \big(-e^{i \delta _{14}} c_{13} c_{24} s_{14} s_{34}-e^{i \delta _{13}}
s_{13} \big(c_{23} c_{34} \nonumber \\
&&-e^{i \delta _{24}} s_{23} s_{24} s_{34}\big)\big)\big)+e^{- i \alpha } m_2 \big(c_{12} c_{23} c_{24}+s_{12} \big(-e^{i
\delta _{13}} c_{24} s_{13} s_{23}-e^{i \big(\delta _{14}-\delta _{24}\big)} c_{13} s_{14} s_{24}\big)\big) \big(-c_{12} \big(c_{34} s_{23} \nonumber \\
&&+e^{i
\delta _{24}} c_{23} s_{24} s_{34}\big)+s_{12} \big(-e^{i \delta _{14}} c_{13} c_{24} s_{14} s_{34}-e^{i \delta _{13}} s_{13} \big(c_{23} c_{34}-e^{i
\delta _{24}} s_{23} s_{24} s_{34}\big)\big)\big)
\end{eqnarray}
\begin{eqnarray}
M_{\tau \tau} &=& e^{ i \big(-\gamma + 2 \delta_{14}\big)} c_{14}^2 c_{24}^2 m_4 s_{34}^2+e^{ i \big(-\beta +2\delta_{13}\big)} m_3 \big(e^{-i \big(\delta
_{13}-\delta _{14}\big)} c_{24} s_{13} s_{14} s_{34}+c_{13} \big(-c_{23} c_{34}+e^{i \delta _{24}} s_{23} s_{24} s_{34}\big)\big){}^2 \nonumber \\
&& +m_1\big(s_{12} \big(c_{34} s_{23}+e^{i \delta _{24}} c_{23} s_{24} s_{34}\big)+c_{12} \big(-e^{i \delta _{14}} c_{13} c_{24} s_{14} s_{34}-e^{i
\delta _{13}} s_{13} \big(c_{23} c_{34}-e^{i \delta _{24}} s_{23} s_{24} s_{34}\big)\big)\big){}^2 \nonumber \\
&& +e^{- i \alpha } m_2 \big(c_{12} \big(c_{34}
s_{23}+e^{i \delta _{24}} c_{23} s_{24} s_{34}\big)-s_{12} \big(-e^{i \delta _{14}} c_{13} c_{24} s_{14} s_{34}-e^{i \delta _{13}} s_{13} \big(c_{23}
c_{34} -e^{i \delta _{24}} s_{23} s_{24} s_{34}\big)\big)\big){}^2
\end{eqnarray}
\begin{eqnarray}
M_{es} &=& c_{14} \big(e^{i \delta _{14}} c_{24} c_{34} \big(e^{- i \gamma } m_4-e^{- i \alpha } c_{13}^2 m_2 s_{12}^2-e^{- i \beta } m_3 s_{13}^2\big)
s_{14}-e^{i \delta _{13}} c_{13} \big(e^{- i \beta } m_3-e^{- i \alpha } m_2 s_{12}^2\big) s_{13}  \nonumber \\
&& \big(e^{i \delta _{24}} c_{34} s_{23} s_{24}+c_{23}
s_{34}\big)+c_{12} c_{13} \big(m_1-e^{- i \alpha } m_2\big) s_{12} \big(e^{i \delta _{24}} c_{23} c_{34} s_{24}-s_{23} s_{34}\big) \nonumber \\
&& -c_{12}^2
c_{13} m_1 \big(e^{i \delta _{14}} c_{13} c_{24} c_{34} s_{14}-e^{i \delta _{13}} s_{13} \big(e^{i \delta _{24}} c_{34} s_{23} s_{24}+c_{23} s_{34}\big)\big)\big)
\end{eqnarray}
\begin{eqnarray}
M_{\mu s} &=& e^{i \big(2 \gamma +2 \delta _{14}-\delta _{24}\big)} c_{14}^2 c_{24} c_{34} m_4 s_{24}+e^{i \big(2 \beta +\delta _{13}\big)} m_3 \big(e^{i
\delta _{13}} c_{13} c_{24} s_{23}-e^{i \big(\delta _{14}-\delta _{24}\big)} s_{13} s_{14} s_{24}\big) \nonumber \\
&& \big(-e^{-i \big(\delta _{13}-\delta
_{14}\big)} c_{24} c_{34} s_{13} s_{14}-c_{13} \big(e^{i \delta _{24}} c_{34} s_{23} s_{24}+c_{23} s_{34}\big)\big)+m_1 \big(-c_{23} c_{24}
s_{12}+c_{12} \big(-e^{i \delta _{13}} c_{24} s_{13} s_{23} \nonumber \\
&&  -e^{i \big(\delta _{14}-\delta _{24}\big)} c_{13} s_{14} s_{24}\big)\big)\big(s_{12}
\big(e^{i \delta _{24}} c_{23} c_{34} s_{24}-s_{23} s_{34}\big)+c_{12} \big(-e^{i \delta _{14}} c_{13} c_{24} c_{34} s_{14}+e^{i \delta _{13}}
s_{13} \big(e^{i \delta _{24}} c_{34} s_{23} s_{24} \nonumber \\
&& +c_{23} s_{34}\big)\big)\big)+e^{- i \alpha } m_2 \big(c_{12} c_{23} c_{24}+s_{12}\big(-e^{i
\delta _{13}} c_{24} s_{13} s_{23}-e^{i \big(\delta _{14}-\delta _{24}\big)} c_{13} s_{14} s_{24}\big)\big) \big(c_{12} \big(-e^{i \delta
_{24}} c_{23} c_{34} s_{24} \nonumber \\
&&+s_{23} s_{34}\big)+s_{12} \big(-e^{i \delta _{14}} c_{13} c_{24} c_{34} s_{14}+e^{i \delta _{13}} s_{13} \big(e^{i
\delta _{24}} c_{34} s_{23} s_{24}+c_{23} s_{34}\big)\big)\big)
\end{eqnarray}
\begin{eqnarray}
M_{\tau s} &=& e^{ i \big(-\gamma +2\delta_{14}\big)} c_{14}^2 c_{24}^2 c_{34} m_4 s_{34}+e^{ i \big(-\beta +2 \delta_{13}\big)} m_3 \big(-e^{-i \big(\delta
_{13}-\delta _{14}\big)} c_{24} c_{34} s_{13} s_{14}-c_{13} \big(e^{i \delta _{24}} c_{34} s_{23} s_{24}+c_{23} s_{34}\big)\big)  \nonumber \\
&& \big(-e^{-i
\big(\delta _{13}-\delta _{14}\big)} c_{24} s_{13} s_{14} s_{34}+c_{13} \big(c_{23} c_{34}-e^{i \delta _{24}} s_{23} s_{24} s_{34}\big)\big)+m_1
\big(s_{12} \big(e^{i \delta _{24}} c_{23} c_{34} s_{24}-s_{23} s_{34}\big) \nonumber \\
&& +c_{12} \big(-e^{i \delta _{14}} c_{13} c_{24} c_{34} s_{14}+e^{i
\delta _{13}} s_{13} \big(e^{i \delta _{24}} c_{34} s_{23} s_{24}+c_{23} s_{34}\big)\big)\big) \big(s_{12} \big(c_{34} s_{23}+e^{i \delta
_{24}} c_{23} s_{24} s_{34}\big) \nonumber \\
&& +c_{12} \big(-e^{i \delta _{14}} c_{13} c_{24} s_{14} s_{34}-e^{i \delta _{13}} s_{13} \big(c_{23} c_{34}-e^{i
\delta _{24}} s_{23} s_{24} s_{34}\big)\big)\big)+e^{- i \alpha } m_2 \big(c_{12} \big(-e^{i \delta _{24}} c_{23} c_{34} s_{24}+s_{23} s_{34}\big) \nonumber \\
&& +s_{12}
\big(-e^{i \delta _{14}} c_{13} c_{24} c_{34} s_{14}+e^{i \delta _{13}} s_{13} \big(e^{i \delta _{24}} c_{34} s_{23} s_{24}+c_{23} s_{34}\big)\big)\big)
\big(-c_{12} \big(c_{34} s_{23}+e^{i \delta _{24}} c_{23} s_{24} s_{34}\big) \nonumber \\
&& +s_{12} \big(-e^{i \delta _{14}} c_{13} c_{24} s_{14} s_{34} -e^{i
\delta _{13}} s_{13} \big(c_{23} c_{34}-e^{i \delta _{24}} s_{23} s_{24} s_{34}\big)\big)\big)
\end{eqnarray}
\begin{eqnarray}
M_{ss} &=& e^{- i \big(\gamma +\delta _{14}\big)} c_{14}^2 c_{24}^2 c_{34}^2 m_4+e^{ i \big(-\beta + 2 \delta_{13}\big)} m_3 \big(e^{-i \big(\delta
_{13}-\delta _{14}\big)} c_{24} c_{34} s_{13} s_{14}+c_{13} \big(e^{i \delta _{24}} c_{34} s_{23} s_{24}+c_{23} s_{34}\big)\big){}^2 \nonumber \\
&& +m_1 \big(s_{12}
\big(e^{i \delta _{24}} c_{23} c_{34} s_{24}-s_{23} s_{34}\big)+c_{12} \big(-e^{i \delta _{14}} c_{13} c_{24} c_{34} s_{14}+e^{i \delta _{13}}
s_{13} \big(e^{i \delta _{24}} c_{34} s_{23} s_{24} +c_{23} s_{34}\big)\big)\big){}^2 \nonumber \\
&&+e^{- i \alpha } m_2 \big(c_{12} \big(-e^{i \delta _{24}}
c_{23} c_{34} s_{24}+s_{23} s_{34}\big) \\ \nonumber
&& +s_{12} \big(-e^{i \delta _{14}} c_{13} c_{24} c_{34} s_{14}+e^{i \delta _{13}} s_{13} \big(e^{i \delta
_{24}} c_{34} s_{23} s_{24}+c_{23} s_{34}\big)\big)\big){}^2
\end{eqnarray}

\end{widetext}

%%%%%%%%%%%%%%%%%%%%%%%%%%%%%%%%%%%%%%%%%%%%%%%%%%%%%%%%%%%%%%%%%%%%%%%%%%%%%%%%%%%%%

\end{document}